\documentclass[pop,fleqn]{w-art}
\usepackage{times}
\usepackage{w-thm}
\usepackage[]{graphicx}

\def\be{\begin{equation}}
\def\ee{\end{equation}}
\def\ba{\begin{eqnarray}}
\def\ea{\end{eqnarray}}

\def\beqa{\begin{eqnarray}}
\def\eeqa{\end{eqnarray}}
\def\fn{\footnote}
\def\p{\partial}

\def\d{\delta}
\def\p{\partial}

\def\ti{\tilde}

\def\a{\alpha}

\def\sq{\sqrt{1-B\dot T^2}}
\def\cA{{\mathcal{A}}}

\begin{document}

\DOIsuffix{theDOIsuffix}
\Volume{51}
\Issue{1}
\Month{01}
\Year{2003}
\pagespan{3}{}
\Receiveddate{}
\Reviseddate{}
\Accepteddate{}
\Dateposted{}
\begin{flushright}
DAMTP-2006-1
\end{flushright}



\title[Electric Tachyon Inflation]{Electric Tachyon Inflation}


\author[Daniel Cremades]{Daniel Cremades%
  \footnote{\quad E-mail:~\textsf{d.cremades@damtp.cam.ac.uk}
           }}
\address[]{Department of Applied Mathematics and Theoretical Physics. Wilberforce Road, Cambridge, CB3 0WA. United Kingdom}
\begin{abstract}
We propose that under certain conditions the universal open 
string tachyon 
can drive topological inflation in moduli stabilised frameworks. 
Namely, the presence of electric 
field in
the world volume of the D-brane can slow down its decay leading to a 
phenomenological model of inflation. The conditions for inflation to 
take place are difficult to satisfy in the standard warped deformed 
conifold but easier to realise in other geometries.
\end{abstract}
\maketitle                   





\section{The Problem of Inflation in String Theory}

Inflation has become, throughout the years, in the standard paradigm
of modern cosmology. In spite of its several drawbacks, mainly
associated to the absence of a fundamental theory giving support to
the paradigm, its phenomenological success and the simplicity of the
ideas involved in its solution of the flatness and homogeneity
problems have elevated inflation to the status of unique (effective)
description of the physics of the primordial universe. It is then 
desirable to embed this phenomenologically successful paradigm into a
robust and fundamental theoretical framework. String theory is believed 
to be an example of such framework.

The embedding of inflation in string theory has to face one obvious
difficulty. Inflationary models are generically characterised by
scalar fields with very flat potentials. Whereas string
constructions have, upon compactification, lots of scalar fields
(moduli), generically they exhibit steep runaway potentials. It is
necessary then to have some moduli stabilisation mechanism 
at work to prevent
undesirable effects to happen like decompactification of the
internal manifold instead of inflation.

A second difficulty for inflation in string theory is the standard 
$\eta$-problem of N=1 supergravity models, that was revisited
in \cite{KKLMMT}. Generic mechanisms for moduli stabilisation
involve the generation of non-trivial superpotentials by some
combination of fluxes. If the inflaton field happens to belong to
some chiral multiplet of the underlying N=1 supergravity, additional
contributions to the scalar potential might appear yielding
generically large contributions to the mass of the inflaton that
halt inflation. 
An obvious way to overcome this difficulty is the
consideration of some source of inflation that is explicitly
non-supersymmetric and thus unrelated to the underlying N=1
structure of the theory. Non-BPS branes are an example of such
explicitly non-supersymmetric system whose decay is able to produce
inflation. This work is devoted to the study of the conditions under
which the decay of non-BPS produces inflation in generic Type II
configurations.

\section{Non-BPS D-branes and their effective action}
Apart from standard BPS D-branes, Type II string theory has a number
of non-BPS objects in its spectrum named non-BPS D-branes. A non-BPS
Dp-brane is a p+1-dimensional (p odd (even) for Type IIA(B))
hypersurface in which open strings can end. The
lowest lying spectrum of such objects contains a real
tachyonic mode that signals the instability of the object towards decay.
A closely related system is that of the bound state made by a 
Dp-anti-Dp pair on top of each other. In this case, the tachyonic 
mode living in the world volume can be described by a complex field.

The dynamics of the tachyon field $T(x)$ in both cases are codified
in the following low energy theory effective action
\cite{Sen1,GKL}\beqa S=-T_p\int d^{p+1}x
~V(T)\sqrt{g_{\mu\nu}+D_\mu TD_\nu T+(F_{\mu\nu}+B_{\mu\nu})}+\int
dT\wedge W(T)e^{F+B}\wedge C \label{effaction}\eeqa where $T_p$ is
the tension of object, to be specified below; $D$ stands for the
covariant derivative with respect to the corresponding gauge 
field living in the world-volume of the brane\fn{Note 
that in the case of the single non-BPS brane the 
tachyonic mode is uncharged.},
whose field strength is given by $F_{\mu\nu}$; $g_{\mu\nu}$ and
$B_{\mu\nu}$ are the pullbacks of the metric and B-field in the
worldvolume of the brane and $C$ is the standard formal sum of RR
potentials that appear in the CS terms. $V(T)$ is the tachyon
potential and $W(T)$ is a function whose asymptotic behaviour is
known but whose form will not be relevant in what follows. The
tachyon potential is given by \cite{LLM} \beqa V(T)={1\over
\cosh({T\over \sqrt{2\a'}})}. \label{potential}\eeqa
In spite of the subtleties required to interpret (\ref{effaction})
as an effective action, it seems to capture some non-trivial features
of the decay of the D-brane. In particular, it leads to the correct
energy-momentum tensor as computed by CFT methods (see \cite{sen_review} 
for a review).
Since this is the coupling of the system to the gravitational
field, the action (\ref{effaction}) 
is the correct tool we must use to describe
tachyon decay is cosmological settings, where we are mostly
interested in the classical coupling between the source of inflation
and the gravitational field, rather than in details concerning the
quantum behaviour of the system at the microscopic level

Even considering the action (\ref{effaction}) together with the
potential at a purely classical level, some peculiarities arise.
First of all, the tachyon condenses at $T\to \infty$ where $\dot
T\to 1$. Secondly, at $T\to \infty$ the action vanishes. This is in
accordance with the fact that the non-BPS D-brane should decay to
something topologically identical to the closed string vacuum, with
no open string degrees of freedom present. One can compute the
density and pressure of the system when $T\to\infty$ finding \beqa
p=M_{pl}^4{V(T)\over\sqrt{1-\dot T^2}}\to{\rm const.},\quad \ \quad
p=-M_{pl}^4V(T)\sqrt{1-\dot T^2}\to 0.\eeqa This pressureless fluid
is conventionally called {\it tachyon matter} \cite{sen_matter}, and
it was given in \cite{LLM} the microscopic description of a coherent
state of highly excited closed string modes
($m\sim 1/g_s$) localised near the original position of the non-BPS
branes. They found that the ratio of the pressure of the system with
respect to its energy density was of order $g_s$, and thus in the
classical limit both descriptions match.

In \cite{2D} was proposed that both descriptions do not only
hold at the classical (tree) level but one can make a complete
correspondence even at the quantum level. The proposal (called the
{\it completeness conjecture}) states that in principle one can
formulate an open string field theory describing the decay of the
D-brane just in terms of open strings, and no explicit coupling to
closed strings is needed in order to make it consistent. The
conjecture thus states that one can describe the set of closed
string fields arising from the decay and its interactions just in
terms of open strings, and this description in terms of open strings
is complete. Remarkably, this conjecture has been proven true in two
dimensional string theory \cite{2D}.

Finally, let us point out one final result that will be useful for
our purposes. It was found in \cite{Partha} that the decay of a
non-BPS system is slowed down by a factor $\sqrt{1-E^2}$ when
electric field (or, analogously, a density of fundamental string
charge) has been turned on in the world volume of the brane. This is
fact will be most useful for our results in what follows.

\section{Tachyon Inflation}
\subsection{Inflation basics}
The proposal of tachyon inflation is to use the energy released by
the D-brane decay to produce inflation\cite{Sen_cosmo,tachyonic}. 
Our starting point is the 
four dimensional action
\be S= \int
\,d^4x\sqrt{-\bar g}\left({M_{pl}^2\over 2} R-\cA
V(T)\sqrt{1+B\partial_\mu T\partial^\mu
T}\right)\,,\label{actioneg1} \ee where $\cA$ and B are,
respectively, dimension four and dimensionless parameters that come
from the dimensional reduction of (\ref{effaction}), as we will see.
Let us assume a standard FRW
ansatz\fn{The obvious question is if such an ansatz is a solution of
the supergravity equations of motion. In general this question is
very difficult to answer since time-dependent backgrounds for string
theory have always been problematic. Here we are taking an effective
field theory approach, in which all the fields that are not relevant
for the physics under consideration are assumed to be integrated
out. For this to be consistent, the energies involved in the low
energy theory have to be smaller than the typical scale associated
to these higher dimensional modes, that is assumed to be the string
scale.} for the metric $d\bar s^2=\bar g_{\mu\nu}dx^\mu dx^\nu$:
\beqa d\bar s^2 = -(dt)^2+a(t)^2 \left({dr^2\over
  1-kr^2}+r^2d\theta^2+\sin^2\theta d\phi^2\right)\,. \label{FRW}
\eeqa The corresponding equations of motion are respectively the
Friedmann (constraint) equation, the evolution equation for the
scale factor $a(t)$ (recall that $H\equiv\dot a/a$) and the equation of
motion for the tachyon field: \ba H^2&=&{1\over 6} {A \cdot V(T)
\over \sqrt{1-B\dot T^2}} -{k\over a^2}\,,
\label{friedmann}\\
{\ddot a\over a}&=& {A \cdot V(T)\over6 \sqrt{1-B\dot T^2}}
\left({1-{3\over 2}B\dot T^2}\right)\label{evolution}\\
3H{AVB\dot T\over \sq}&+&\p_t\left({ABV\dot T\over \sq}
\right)=-A\sq ~V'(T), \label{tachy} \ea
where $A\equiv 2\cA/M_{pl}^2$. In order to solve this system of 
differential equations one needs a set of initial conditions of $a(t)$
and $T(t)$. We will
choose time symmetric initial conditions in order to obtain the
(classical description of the) kind of decay described in
\cite{Sen_cosmo}: this implies $\dot T(0)=\dot a(0)=0$.
Automatically this implies $k=1$ to satisfy Friedmann's equation and
consequently $a(0)=(6/AV(T_0))^{1/2}$. The solution is then
completely determined in terms of $A$, $T_0$ and $B$. A complete analysis
of this system of equations\fn{This analysis also includes the
coupling of the system to the volume modulus of the compactification
manifold, under the assumption that its dynamics are governed by a 
KKLT-like potential, in order to examine
how cosmological inflation can be compatible with moduli stabilisation.}
has been performed in \cite{CQS}. This analysis 
shows that the evolution of the system takes place mainly 
in the slow-roll regime; we will therefore assume these conditions
from here on.

In order to connect with the standard inflationary terminology, we
have to define a canonically normalised inflaton field in terms of
the tachyon as
\be \phi \ =\ \sqrt{\cA B}\int \sqrt{V(T)}\ dT \label{canon} \ee
As in standard inflation, we will assume that $T$ is a spatially
homogeneous, time dependent field. The slow-roll parameters measure
the ability of some potential to produce inflation. The form of
these parameters in terms of $T$ is\fn{Note that the functional
form of the $\eta$ parameter makes the system suitable for an
$\eta$-problem if the coefficient of cosh $x$ is of order one.} \be
\epsilon = \frac{M_{pl}^2}{4\alpha' \cA B}\ \frac{\sinh^2 x}{\cosh
x}, \qquad \eta= -\frac{M_{pl}^2}{2\alpha' \cA B}\ \cosh
x\left[1-\frac{3}{2}\, {\rm {tanh}}^2 x\right], \label{eta}\ee with
$x\equiv \left(\frac{T}{\sqrt{2\alpha'}}\right)$, and both of them
must be much smaller than one to yield a phenomenologically
acceptable inflationary scenario. The number of e-folds in the slow
roll approximation is then given by 

\be N\simeq \frac{\cA
B}{M_{pl}^2}\ \int_{T_e}^{T_b} {V^2\over V'}\, dT\, =
\frac{2\alpha' \cA B}{M_{pl}^2} \
\log\left[\coth\left(\frac{T_b}{\sqrt{8\alpha'}}\right)\right],\label{efolds}\ee 
where
' means now derivative with respect to $T$, and $T_b$ and $T_e=\infty$ are,
respectively, the values of the tachyon field at the beginning and
end of inflation. The
amplitude of density perturbations is given by \be \delta_H \simeq\
\frac{\sqrt{\cA^2 B}}{5\pi \sqrt{3} M_{pl}^3}\,
  \frac{V^2}{V'}=\sqrt{\frac{\alpha'\cA^2 B}{75\pi^2 M_{pl}^6}}
\, { \rm csch }\, \left(\frac{T_b}{\sqrt{2\alpha'}}\right). \ee 

As it is well known, to match observational constraints 
it is necessary to have $N\geq 65$,
$\d_H\simeq 2\times 10^{-5}$. This last numerical value is given by the
COBE normalisation, which must be computed at the point when the last 60 
e-folds of inflation start.
Here
we can already see what is the main phenomenological point of
tachyon inflation. Assuming that generically we will have
$\a'/M_{pl}^2\leq 1$, if one manages to get a value of the
adimensional quantities \beqa {\a'\cA\over M_{pl}^2}B\gg 1, \qquad \
\qquad \left({\a'\cA\over M_{pl}^2}\right)^2B\ll1,\eeqa then
generically we will have phenomenologically successful inflation,
what basically amounts for a small value of $\a'\cA/ M_{pl}^2$ and a
large value of $B$.

\subsection{Compactification of the tachyon action}
Let us see now what is the origin of these parameters $\cA$ and $B$ that 
appeared in the four dimensional effective action (\ref{effaction}).
Consider a $p+1$-dimensional non-BPS system in Type II theory\fn{We
prefer to maintain the discussion generic at this point. In Type
IIA(B), the object is a single non-BPS brane for $p$ odd (even) and
a pair D$p$-anti-D$p$ for $p$ even (odd).}, wrapping the whole four
dimensional space-time and some $p-3$-cycle in the internal space.
This system will be described by the tachyon action
(\ref{effaction}). Consider the embedding of this action on a metric
of the form \be ds_{10}^2=e^{2\omega(y)}\bar g_{\mu\nu}dx^\mu dx^\nu
+ \ti g_{mn} dy^mdy^n \label{metric_gral}\ee and its dimensional
reduction down to four dimensions. The six-dimensional metric $\ti
g$ has the generalised Calabi-Yau form that is found in generic flux
compactifications, and whose moduli are 
supposed\fn{An explicit study of inflation in moduli stabilised
frameworks requires knowledge of the explicit stabilisation mechanism. 
It was performed in \cite{CQS} a complete analysis of tachyon inflation 
in the presence of moduli stabilisation for a particular IIB model
with a single Kahler modulus, subject to a KKLT-like potential. We 
assume that other moduli stabilisation mechanisms can be 
analysed and compatibilised with tachyon inflation in a similar way.}
to be
stabilised\fn{As emphasised in section 1, this can turn out to be a
non trivial constraint. Even if in our case the tachyon field will
not belong to any kind of chiral multiplet and thus in principle its
dynamics will not be (directly) influenced by the particular
mechanism of moduli stabilisation taking place, as it was the case
in \cite{KKLMMT}, generic moduli potentials can put non trivial
upper bounds on the value of the dimensionful parameter $\cA$ in
order to prevent undesirable effects in cosmological settings like
decompactification of the internal dimensions.}. The reduction down
to four dimensions can in principle be very complicated but we can
make some simplifying assumptions. First, we will assume that the
warping factor $\omega(y)$ is constant along the directions of the
brane. Secondly, we will suppose that the pullback of the
$B$-field and the gauge field are zero in the world-volume of the
brane (we will relax this condition later). Under these assumptions,
the four dimensional action takes the form (\ref{actioneg1}) 
with\fn{In the absence of internal magnetic field, the lowest 
tachyonic mode has a constant profile over the internal dimensions 
and thus the CS term will not contribute to the effective four 
dimensional action.}
\beqa \cA=e^{4\omega}T_p V_{p-3}\, , \ \qquad \ \qquad
B=e^{-2\omega}, \eeqa with \beqa V_{p-3}=\int d^{p-3}y \ \sqrt{P[\ti
g]_{ab}}\, , \ \qquad \ \qquad T_p=\sqrt{2}
{(2\pi)^{-p}\a'^{-{{p+1}\over 2}}\over g_s},\eeqa where $P[\ti g]$
is the pullback of the internal metric in the world volume of the
brane. Note that the tension of the $p$-brane gets an extra
$\sqrt{2}$ factor in the case of the brane-antibrane pair.

The scales involved in the problem are as follows. The Planck mass
is given by \beqa {M_{pl}^2\over 2}={(2\pi)^{-7}\over g_s^2
\a'^4}\int d^6y\sqrt{\ti g} \ e^{2\omega}.\eeqa The (closed) string
scale is not known since in general it is not possible to quantise
the string in arbitrary backgrounds. However, we can estimate it as
some numerical value $c$ that is given e.g. by the fluxes in the
configuration times $\a'$. The (open) string scale in a warped
compactification will depend in the point in the internal manifold
where the subset of open strings under consideration is attached. In
our case it will be set by the mass of the tachyon and thus will be
given by $m_s^2\sim {c\over B\a'}$. As we will see, under the
circumstances of interest for our problem we will be able to set
$c=1$. The Hubble scale is the typical energy scale during inflation
and is given by $H^2=\cA V/(3 M_{pl}^2)$. For our approximation to
make sense, then, the Hubble scale must be smaller than the string
scale at all times. But this seems to lead to a contradiction, since
comparison with (\ref{eta}) and (\ref{efolds}) implies that either
$N$ is small or $\eta$ is of order 1. There are two possible ways to
overcome this difficulty. First, to assume that $c$ has a value
somewhat bigger than one. Note that for $c\sim O(10)$ the string
scale would be twice as large as the Hubble scale. Secondly, the
addition of electric field in the world-volume of the brane will
lower the Hubble scale keeping the string scale fixed. Let us review
how the addition of world-volume electric field will generically
help in obtaining phenomenologically successful tachyon inflation.

\subsection{Addition of electric field}
As advanced, it is known that the addition of electric field in the
world volume of a non-BPS D-brane can slow down its decay. We want
to make use of this fact to generate long-lasting inflation, and we
will see that the consideration of this degree of freedom is also
convenient for other phenomenological purposes. To turn on electric
field along one compact direction in the world volume of the D-brane
it is necessary for the cycle wrapped by the D-brane to possess
non-trivial one cycles\fn{Note that one can consider one-cycles that
are trivial on a Calabi-Yau but non-trivial when restricted to some
higher dimensional $p-$cycle, $p<6$.}. We will generate this electric 
field by turning on a world-volume gauge potential that has (locally)
the form $A=-Exdt$, where $x$ is the coordinate that parametrises 
the 1-cycle. This vector potential cannot be transformed into a 
time-dependent Wilson line because of the topology of the space and thus
the presence of this constant electric field is consistent with the
symmetries of the system and the cosmological evolution, provided
it satisfies the Maxwell equation
\beqa
\partial_{x}{\sqrt{-\bar g}\cA V(T)E\over \sqrt{1-B\dot T^2-E^2}}=0.
\eeqa 
This tells us that if $\bar g$, $\cA$ and $B$ are independent
of the internal coordinate $x$ then a constant $E$ will satisfy this
equation. This is true for example when the $p-3$ cycle wrapped by
the D-brane is a torus. In this case the 4-d effective Lagrangian
can be rewritten as \be \hat A V(T) \sqrt{-\bar g}\sqrt{1-\hat B
\dot T^2}\,, \label{electric} \ee where $\hat A=A \sqrt{1-E^2}$ and
$\hat B=B/(1-E^2)$. Thus the total number of e-folds, which is
proportional to $AB$, would increase by a factor $1/\sqrt{1-E^2}$,
whereas the density perturbations will remain 
constant\fn{Pushing the value of $E$ to be close to 1 should not
be considered as fine-tuning, since the relevant physical variable
is not $E$ but the electric charge $q\sim E/\sqrt{1-E^2}$ 
\cite{Gibbons}. Putting $E$ close to 1 is just equivalent to 
increasing $q$.}. Note that
the addition of electric field will also help in the problem with
the scales mentioned in the previous section, since it will lower
the Hubble scale keeping constant the string scale (which is
unaffected by the presence of electric field). Also, the electric
field induces in general fundamental string charge 
in the world volume of the
brane \cite{Gibbons}, and thus these fundamental strings will be 
left over as a
remnant of the tachyon condensation. Note, however, that in a
manifold without one cycles like a CY these fundamental strings will
generically contract to zero size and annihilate into the vacuum, so
that the presence of F-cosmic strings is not expected in these
situations.

In order to illustrate how the presence of electric field can help 
obtaining phenomenological numbers for inflation, let us consider the 
sample values for the parameters 
(in units in which $M_{pl}^2=1$) $A=10^{-6}$,
$B=5 \cdot 10^9$, $T_b=0.1$, $\a'=10^{-3}$. Then the number of e-folds is
$N=1.07$ and the density perturbations are given by 
$\d_H=1.2\cdot 10^{-5}$. Turning on 60 units of electric field 
($E/\sqrt{1-E^2}=60$) one gets $N=64.38$ and $\d_H$ remains unaltered.
With respect to the scales, one has $m_s^2=2\cdot 10^{-7}$ and 
$H^2=6\cdot 10^{-10}$, in units where $M_{pl}^2=1$. This kind of 
estimations are confirmed by the full numerical analysis, see \cite{CQS}
for details.

\subsection{Topological inflation}
The fact that the tachyon potential (for the case of the non-BPS brane) 
has a $\bf{Z}_2$ symmetry $T\to -T$ might lead to the concern that
the generic presence of domain wall defects 
upon tachyon condensation in this kind of potentials could make the 
model completely incompatible with observational constraints. This is not
the case. Actually, this model is a concrete realisation of topological 
inflation in string theory.

Topological inflation was introduced in \cite{andrei} partly
as a way to ameliorate the issue of initial conditions of standard
inflationary models. 
In topological inflation the initial conditions are governed by
the existence of a topological defect at the maximum of a
potential with degenerate minima. If the defect is thicker than
the Hubble size $H^{-1}$, then a patch within the defect  can
inflate in all directions (including the ones transverse to the
defect) and become our universe. The conditions for the thickness
of the defect happen to be satisfied if the slow-roll condition,
requiring $\eta$ to be small, holds. This is easy to see since the
thickness of the defect ($\delta$) is inversely proportional to
the curvature of the potential at the maximum which itself is
proportional to $\eta$. Then,  the condition $\delta\gg H^{-1}$ is
satisfied if $\eta \ll 1$.

\section{Explicit Examples}
The search for explicit examples is the search for appropriate
compactification manifolds.
\subsection{Toroidal compactification}
Even if this is a phenomenologically not very interesting case,
since generically toroidal models have the string scale very close
to the Planck scale, the torus is still interesting since it allows
to appreciate most of the features of the problem. Consider the
non-BPS $p$-brane wrapping $p-3$ of the radial coordinates of the
torus. Evaluation of this trivial case yields automatically \beqa
\cA={\sqrt{2}(2\pi)^{-p}\a'^{-2} V_l\over g_s} \,, \qquad B=1, \,
\qquad {M_{pl}^2\over 2}={V\over (2\pi)^7 g_s^2 \a'},\eeqa times a
$\sqrt{2}$ factor in the case of brane-antibrane pairs, where $V_l$
is the volume of the directions wrapped by the brane and $V$ the
volume of the torus, both in $\a'$ units. The number of e-folds and
density perturbations are given by \beqa
N&=&{(2\pi)^{7-p}g_s\over \sqrt{2} V_t}\\
\d_H^2&=&{(2\pi)^{21-2p}g_s^4\over 75\pi^2 V_l V_t^2},\eeqa with
$V_t$ the transverse volume to the brane in $\a'$ units. One can see
that phenomenological numbers can be obtained e.g. setting $g_s\sim
0.1$, $V_t\sim 1$, $V_l\sim 10^{10}$, although obviously the
addition of electric field can make the situation better. What this
simple model shows us is that (in the absence of electric field)
large volumes transverse to the brane seem to render tachyon
inflation useless, at least as the main source of inflation.

\subsection{Warped Deformed Conifold}
The warped deformed conifold is one of the most celebrated
geometrical constructions in (IIB) string theory since it provides an 
example of a throat
created by the backreaction of fluxes in which the metric is
explicitly known \cite{KS}. Though the metric is known, two limits are
specially relevant for our purposes. The apex of the throat has the 
topology of a $S^3$, and the metric is given by
\beqa
ds^2_7=e^{2\omega}d\bar s_4^2+ \ti R^2 \  ds_{S^3}^2,\eeqa with
$\omega=-2\pi K/(3g_sM)$ and $\ti R^2=g_s M\a'$. Far from the apex the 
geometry looks like $AdS^5\times T^{1,1}$:
\beqa ds_{10}^2={r^2\over R^2}d\bar
s_4^2+{R^2\over r^2}(dr^2+r^2ds_{T^{1,1}}^2)\eeqa with $R^4={27\over
4}\pi g_s MK\a'^2$.
Let us consider the decay of a non-BPS D6 brane wrapped in the $S^3$ in
the apex of the deformed conifold.
The computation of $\cA$ and $B$ yields
\beqa \cA=e^{-{8\pi K\over 3g_sM}}{\sqrt{2}\over
16\pi^3}g_s^{1/2}M^{3/2}, \hspace{1cm} \ B=e^{4\pi K\over
3g_sM},\eeqa
where $M$ and $K$ are the RR and NSNS flux density along 
the 3-sphere and its dual cycle, respectively. The Planck mass is 
given in this case by
\beqa
{M_{pl}^2\over 2}={V_6\over (2\pi)^7 g_s^2 \a'^7}, \hspace{1cm} 
V_6=k\left({2\pi\over 3}\right)^3\left({27\pi\over
4}g_sMK\right)^{3/2},
\eeqa
where $k\geq 1$ is some number. The number of e-folds is given by
\beqa N\simeq {2\a'\cA B\over M_{pl}^2} ={27\over 2}\left({4\over
27\pi}\right)^{3/2}{g_s e^{-4\pi K\over 3 g_s M}\over k
K^{3/2}}\simeq { 10^{-2}  e^{-4\pi K\over 3 g_s M}\over
kK^{3/2}},\eeqa
that is a number at most of order $10^{-2}$. Note that one cannot 
ameliorate this result by the addition of electric field since there are
no 1-cycles in a $S^3$. This result teaches us that in
general warped configurations will not help in obtaining realistic 
models of inflation, and also that in general one is more likely to 
obtain (in the absence of electric field) phenomenological inflation
in situations where the string scale is close to the Planck scale.

\subsection{Other possible constructions}
Addition of electric field generically allows for a phenomenologically
viable model of inflation provided that the submanifold $\Sigma$ 
wrapped by the non-BPS brane (or D-brane-anti-D-brane pairs) has 
$b_1(\Sigma)> 0$. One example of such manifolds are special 
Lagrangian manifolds. Explicit examples of SLAGs with $b_1(\Sigma)>0$ 
were provided in \cite{mirror}. An explicit computation of the number 
of e-folds and density perturbations needs the precise knowledge of
the metric, but we have seen that these constraints are met with 
large enough value of the electric charge.

Other kind of constructions in which moduli are stabilised but allow
at the same time for non-trivial one cycles are those involving 
twisted tori\fn{See \cite{Camara} and references therein.}. In the 
particular construction considered in \cite{Camara}, the precise 
computation of the number of e-folds and density perturbations would be
very similar to that of the toroidal case considered in section 4.1.

There are other effects that can ameliorate the phenomenological 
potential of the model like the addition of magnetic fields in the world
volume of the brane. The presence of magnetic fields will generically
improve the number of e-folds in a given configuration, but they will
also complicate the dimensional reduction of the tachyon action. 
In principle they will generate non-trivial profiles for the tachyon
wave function in the internal dimensions, as well as increase the 
number of four-dimensional tachyon fields. A complete investigation
of magnetic tachyon inflation is thus beyond the scope of this work.

\section{Conclusions}
Tachyon inflation provides an example of topological inflation 
within the framework of string theory. The fact that the tachyon 
is a generic object in string theory and the fact that its
presence is always linked to explicitly non-supersymmetric systems (and
therefore free from the problems pointed out in \cite{KKLMMT}) makes it
specially attractive for considering it as a source for inflation.
We have considered here the conditions under which this kind of 
inflation can take place, from an effective field theory point of view.
We have found two main results: that the systems in which it can work 
(in the absence of electric field) are
those in which the string scale is very close to the Planck scale,
and that the presence of electric field in the world-volume of the brane
(or brane-antibrane pairs) that support the tachyon field can greatly 
improve the phenomenological potential of the model.

The end point of the decay is the least understood from a theoretical 
point of view, since most of the classical expectations fail 
when reaching this point \cite{Sen_otro}. A complete understanding 
of this regime will be crucial to address several other interesting 
questions regarding this model like reheating after inflation (that 
could take place along the lines suggested in \cite{cline}) or the 
possible over-closure of the universe due to tachyon matter pointed out
in \cite{kofman}, based in arguments extrapolated from the classical
description of the decay. We leave these questions for future work.

\begin{acknowledgement}
I thank Fernando Quevedo and Aninda Sinha for the collaboration 
in the original paper and enlightening conversations and 
Fernando Marchesano and
Angel Uranga for very useful discussions. I
also thank the organisers of the Corfu 2005 Summer Institute for
the invitation to present this contribution. 
The author's work is supported 
by the University of Cambridge.
\end{acknowledgement}


\end{document}